\documentclass[%
 reprint,
 amsmath,amssymb,
 aps,
]{revtex4-2}

\usepackage{graphicx}% Include figure files
\usepackage{dcolumn}% Align table columns on decimal point
\usepackage{bm}% bold math
\newcommand{\ps}{\mathrm{ps}}

\begin{document}

\preprint{Arxiv/123-QED}

\title{Near-field Ultrafast Nondiffractive Laser Nanoscale Cutting of Glass with Quasi-null Interface Roughness}% Force line breaks with \\

\author{Rajeev Dwivedi}
 \altaffiliation{Laboratoire Hubert Curien, UMR CNRS 5516,Universit\'{e} Jean Monnet, 42000 Saint Etienne, France.}%Lines break automatically or can be forced with \\
\author{Anton Rudenko}%
\altaffiliation{Laboratoire Hubert Curien, UMR CNRS 5516,Universit\'{e} Jean Monnet, 42000 Saint Etienne, France.}%Lines break automatically or can be forced with \\
\author{Guodong Zhang}%
\altaffiliation{School of Artificial Intelligence, Optics and Electronics, Northwestern Polytechnical University, Xi'an 710072, China}%Lines break automatically or can be forced with \\
\author{Xxx Sedao}%
\altaffiliation{Laboratoire Hubert Curien, UMR CNRS 5516,Universit\'{e} Jean Monnet, 42000 Saint Etienne, France.}%Lines break automatically or can be forced with \\
\author{Yaya Lefkir}%
\altaffiliation{Laboratoire Hubert Curien, UMR CNRS 5516,Universit\'{e} Jean Monnet, 42000 Saint Etienne, France.}%Lines break automatically or can be forced with \\
\author{Ciro D'Amico}%
\altaffiliation{Laboratoire Hubert Curien, UMR CNRS 5516,Universit\'{e} Jean Monnet, 42000 Saint Etienne, France.}%Lines break automatically or can be forced with \\
\author{Guanghua Cheng}%
\altaffiliation{School of Artificial Intelligence, Optics and Electronics, Northwestern Polytechnical University, Xi'an 710072, China}%Lines break automatically or can be forced with \\
\author{Razvan Stoian}%
\altaffiliation{Laboratoire Hubert Curien, UMR CNRS 5516,Universit\'{e} Jean Monnet, 42000 Saint Etienne, France.}%Lines break automatically or can be forced with \\
 \email{razvan.stoian@univ-st-etienne.fr}

\date{\today}% It is always \today, today,

\begin{abstract}
Generally, ultrafast laser glass dicing is governed by explosive material removal, which restricts the achievable surface roughness. Here we report single-step dicing of aluminosilicate glass with residual interface roughness below 5\, nm, producing optical-quality surfaces without post-processing, using near-field-enhanced 0th-order Bessel–Gauss beams. This exceptional interface quality cannot be explained by a conventional laser ablation mechanism, such as phase explosion, melt expulsion, or fracture-dominated material removal. Instead, quantitative analysis of local thermal dynamics indicates that material removal is consistent with a Hertz–Knudsen-type evaporation process, enabling layer-by-layer removal with precision approaching a few molecular layers. These results suggest that nanoscale optical confinement fundamentally alters the response of transparent materials to ultrafast laser irradiation and establish a route toward deterministic laser processing with near-atomic precision. 
\end{abstract}

%Use showkeys class option if keyword
\maketitle

%\tableofcontents

\section{\label{sec1}Introduction}
The roughness of diced glass interfaces critically limits performance in optical, microfluidic, and photonic systems \cite{nanofludic, photonicssystemFAN20088, fibre-to-chip-waveguide}, where microscopic imperfections introduce scattering, fluid distortions, and reduce interfacial reliability. Additional complex processing steps are then required. Achieving in-volume glass cutting with roughness comparable to that of optical polishing without post-processing remains a central challenge in precision glass fabrication. This requires a twofold development; a cutting tool addressing in a single step the thickness of the material and a damage-affected zone restricted to nanometer. Such a possibility also raises fundamental questions, pinpointing material-removal processes capable of achieving monolayer resolution with ultrashort lasers. The use of ultrafast laser tools for cutting transparent materials has significantly advanced the field of material dicing, particularly with the advent of non-diffractive Bessel-Gauss beams. Developed to boost the efficiency \cite{malinauskas2016ultrafast}, the technique relies on localized nonlinear energy deposition through the glass generating oriented stress fields favoring cracking and separation \cite{balage2023bessel,liu2024high, dudutis2025polarization}, bypassing the uncontrolled fracture associated with conventional methods \cite{Dudutis:20,meyer2017submicron,sugioka2014ultrafast,Jenne:20}. The spatial and temporal core engineering amplifies the optical energy deposition \cite{Bhuyan2014,chen2019}, and determines preferential relaxation directions with thermal gradients and stress levels in the GPa range, improving cutting uniformity \cite{ungaro2021using,shin2020strength,bhuyan2015high,liu2023engraving}.  The typical roughness reported in laser stress-induced dicing lies above $\mathrm100$\,nm \cite{liu2024high} which reflects the modulation introduced by the optical damage, namely the intrinsic stochasticity of a fracture mechanisms.

Recently, new developments in laser nanostructuring revealed the importance of self-induced near-field enhancement to obtain a volume scribing precision of nanometer level \cite{li2024super,yan2022near}. These evanescent fields arise from the interaction of the incident far field with self-generated nanoscale features, leading to highly localized energy concentration. Combined with the extended propagation of non-diffractive beams, this mechanism has recently enabled ultrahigh-aspect-ratio nanostructuring with lateral dimensions approaching $\lambda/100$ over depths exceeding $100~\mu$m \cite{Zhang25}. Such extreme optical confinement offers an opportunity not only to improve spatial resolution, but also to explore material-removal regimes inaccessible under conventional ultrafast laser irradiation.

At these extreme nanometer scales, the interaction volume is reduced to dimensions comparable to those governing elementary phase transformations, raising the question of whether the mechanisms responsible for ultrafast laser ablation remain unchanged. While ultrafast ablation is generally interpreted as highly nonequilibrium processes involving melting, fracture, or explosive material removal, nanoscale confinement may fundamentally alter the thermodynamic pathways available to the material. Determining the local thermodynamic conditions during processing and identifying the mechanism responsible for nanometer-scale interface regression are, therefore, essential to understanding the ultimate limits of ultrafast laser machining.

We demonstrate that near-field-enhanced non-diffractive ultrafast laser irradiation enables glass separation with nanometer residual roughness, i.e. at optical quality. A roughness result of less than 5\,nm is obtained, directly from an ablative process, substantially lower than that of the existing benchmark by a factor of $50$. It is to be noted that such roughness was observed in nanoscale stealth dicing experiments after wet-etch post-processing \cite{li2024super} and considered a potential attribute of the near-field interaction. The prospect of using nanoablation from near-field confinement along the axis of far-field radiation underlines a laser glass separation process which is not mechanically driven, but a more gentle thermally-driven material separation that approaches atomic resolution. Using time-resolved quantitative imaging we indicate nanoscale domains with temperatures above 3000\,K for tens of ns. In such conditions nucleated phase transitions is dimensionally prohibited. Combined with a theoretical framework that evaluates the interfacial temperature during laser scanning, our analysis reveals that the thermal ablation process relies essentially on a Hertz-Knudsen process of evaporation, capable of nanometer interface regression. The process is optimized for scanning pitch, polarization and pulse duration to provide sub-10\,nm scribing resolution over a depth of more than $\mathrm{100}$\,$\mu$m, directly providing high-quality glass cut surfaces.

\begin{figure*}
\centerline{\includegraphics[width=1\linewidth]{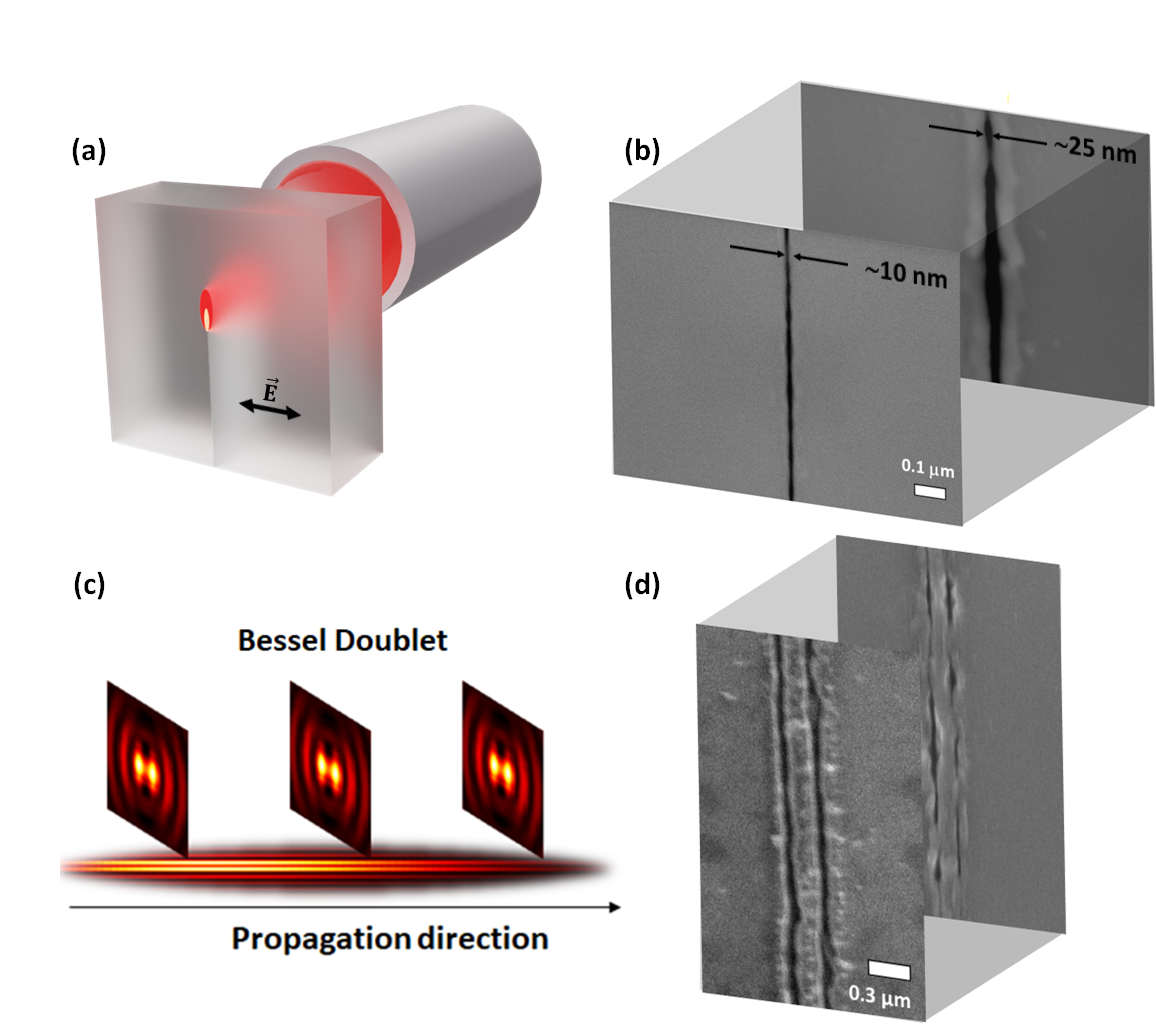}}
\caption{(a) Schematic representation of nanodicing with energy confinement in high aspect ratio-geometries. (b) Electronic microscopy image of front and back surface nanolines obtained in AS87 glass, underlining the extreme energy localisation. A single continuous nanoline is formed at an energy of  $\mathrm{110 \ \%  \ E_{th}}$ (observable scribing threshold). Comparable laser-induced modifications are generated on both the front and rear surfaces of the glass and through the glass. Similarly, multi-Bessel-Gauss beam profiles in (c) fabricate parallel (d) nanotrenches in a 100 $\mu$m thick sample. The irradiation conditions are as follows: pulse duration 200\,fs, scan speed 0.1\,mm/s, repetition rate 100\,kHz. \label{Fig1}}
\end{figure*}

\section{\label{sec2}Results}
\subsection{Near-field-enhanced nanotrenches}
Figure ~\ref{Fig1} sketches the experimental arrangement (Fig. ~\ref{Fig1} (a)) and delivers a first glimpse into the cut characteristics (Fig. ~\ref{Fig1} (b)). The laser passage through the surface, at a local fluence above the damage threshold, $\mathrm{E_{th}}$ (discussed in Supporting Information), opens up a scribed line of a few tens of nanometers in transverse section. A similar track of about 10\,nm appears on the back surface of the glass layer, suggesting a uniform penetration through the glass. The section mismatch with a narrower value on the back surface is the consequence of pure cavitation threshold triggered by a slightly self-focused beam compared with a surface interaction that mixes cavitation and surface ablation.  The transverse section of 10\,nm is more than 100$\times$ smaller than the waist of the laser beam and equals a hundredth of the incident wavelength. The process \cite{Zhang25} is driven by near-field scattering and enhancement on prior nanovoid structures \cite{Bhu10,Bhu17}, developing perpendicular to the laser field at pitch in the nanometer range. The non-diffractive incident beam works as near-field-enhancing cylindrical structure through the whole thickness, producing anisotropic energy deposition and directional ablation. Under continued irradiation and slow scanning, this process evolves into a continuous nanometer-scale trench ((Fig. ~\ref{Fig1} (b)).

\begin{figure*}
\centerline{\includegraphics[width=1\linewidth]{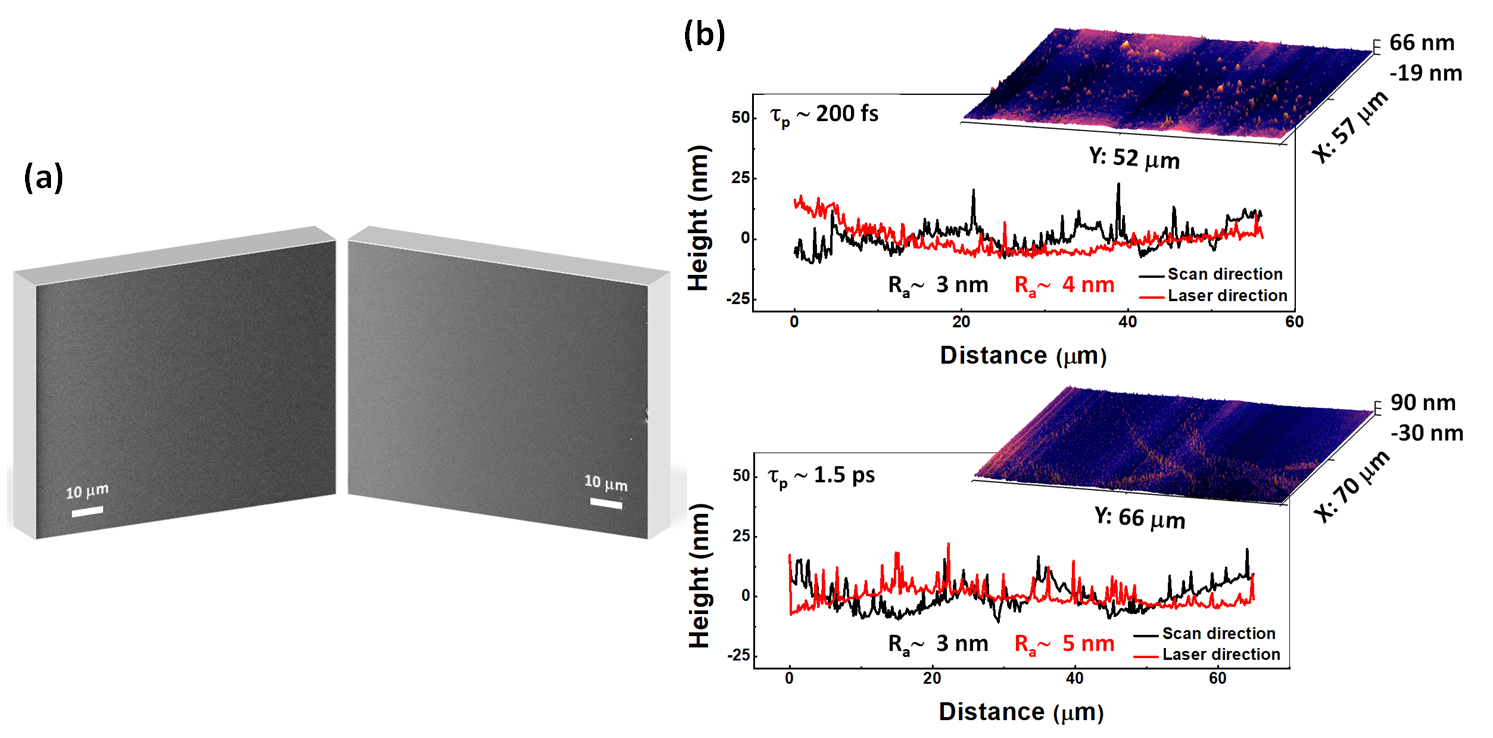}}
\caption{(a) SEM images of the separated interfaces by fs laser pulses. The images demonstrate uniform, smooth cut edges and interfaces over lateral lengths of several tens of micrometer (b)  Three-dimensional AFM topography of a separated surface, showing the overall surface height distribution after cutting with a fs and ps Bessel-Gauss pulse and corresponding AFM line profile extracted along the cutting direction, used to quantify the surface roughness. They yield an average roughness of $\mathrm{R_{a} \ \sim \ 4}$\,nm and  $\mathrm{R_{a} \ \sim \ 5}$\,nm respectively, when measured along laser passage, while remain $\mathrm{R_{a} \ \sim \ 3}$\,nm along the scan direction for both pulse durations (see Supporting Information). \label{Fig2}}
\end{figure*}

The pitch, the distance between two consecutive pulses, in the nanometer range, ensures that the incoming pulses, incident on a void structure, will only deposit energy through scattering and field enhancement on the non-processed part. Decreasing the scan pitch from $\mathrm{10}$\,nm to $\mathrm{1}$\,nm leads to a substantial reduction in trench width, from a few tens of nanometers down to approximately  $\mathrm{6}$\,nm, corresponding to about $\mathrm{20}$ atomic layers in glass. This reflects the ablative part of the near-field distribution. At a fixed nanometer pitch, maintaining the fluence near $\mathrm{110 \ \% \ E_{th}}$ enables the formation of a single, well-defined nanoline on the rear surface of the glass. Extending the pulse duration into the $\ps$ regime is expected to reduce plasma-induced defocusing and enhance energy deposition at the focal plane. While this promotes higher local temperatures and glass flow effects, it also suppresses the formation of ultranarrow trenches, increasing the widths to a few tens of nanometers. The question is now, what state of roughness the interface presents?
Electron microscopy and atomic force microscopy data on the cut interfaces for fs and ps pulse laser irradiation are presented in Fig. ~\ref{Fig2} with Fig. ~\ref{Fig2} (a) showing the image of the residual surface and Fig. ~\ref{Fig2} (b) the topography measurements for irradiation with two generic pulse durations; the fs pulse (200\,fs) and the ps pulse (1.5\,ps). Quantitative roughness distribution measurements are given in Fig. ~\ref{Fig2}, suggesting an average roughness of a few nanometers (more examples are discussed in the Supporting Information).
Figure ~\ref{Fig1} (c)-(d), demonstrate that this technique can be extended to multi-Bessel-Gauss beams (see Supporting Information) to fabricate simultaneous nanotrenches with a separation of $\lambda$/3, demonstrating potential for extracting ultra-thin, high-aspect-ratio nano-sheet with quasi-zero roughness surfaces. 

\subsection{Surface quality of glass cutting}
As observed in Fig. \ref{Fig2}(a), the near-field assisted ultrafast laser cutting separates the glass gently into two pieces, where SEM inspection reveals the absence of brittle fracture in glass (like hackle marks, chipping and crack branching), consequently strengthening the diced edge. The edge profile remains well-defined, indicating that material separation occurs in a controlled manner over the scanning length rather than through catastrophic crack propagation. Subsequently, the AFM topography measurements (Fig. \ref{Fig2}(b)) shows minimal height variation across the scanned area, with an average surface roughness $R_a$ below 5\,nm along the laser passage, while $R_a \sim$ 5\,nm along the scanning direction (see Supporting Information). Height distributions are narrow and spatially uniform, and no periodic surface modulations or localized protrusions are observed. Isolated surface peaks observed in AFM are attributed to extrinsic particulate contamination rather than intrinsic cutting-induced features, indicating that the intrinsic roughness of the laser-treated surface is even lower. These results confirm that the cutting process preserves nanoscale surface integrity without introducing secondary roughening effects. It is nonetheless observable that ps irradiation leads to a higher corrugation of the surface. The ps irradiation usually increases the bulk energy density, triggering processes at the border between thermomechanical effects and phase transition. Energy-dependent processes will be discussed in the next section.

The consistency between SEM and AFM observations suggests that the cutting conditions effectively suppress both surface and subsurface macroscopic catastrophic damage. Under these conditions, the energy remains highly localized, thereby preventing the formation of micrometer scale fracture-induced roughness, even though a fracture-based cleaving will show a reduced roughness on restricted areas. As a result, material removal proceeds through a highly localized, near-field-enhanced process, producing a surface that approaches the smoothness of polished glass. Notably, the low roughness is maintained over extended cutting lengths, indicating good process stability and reproducibility. In contrast to conventional mechanical or laser-based cutting methods, which often require subsequent polishing or etching to reduce surface roughness, the present approach yields nanometer-scale smoothness directly after cutting. This capability significantly simplifies fabrication workflows and reduces the risk of introducing additional defects.

\subsection{Mapping thermodynamic states}
\begin{figure*}
\centerline{\includegraphics[width=\linewidth]{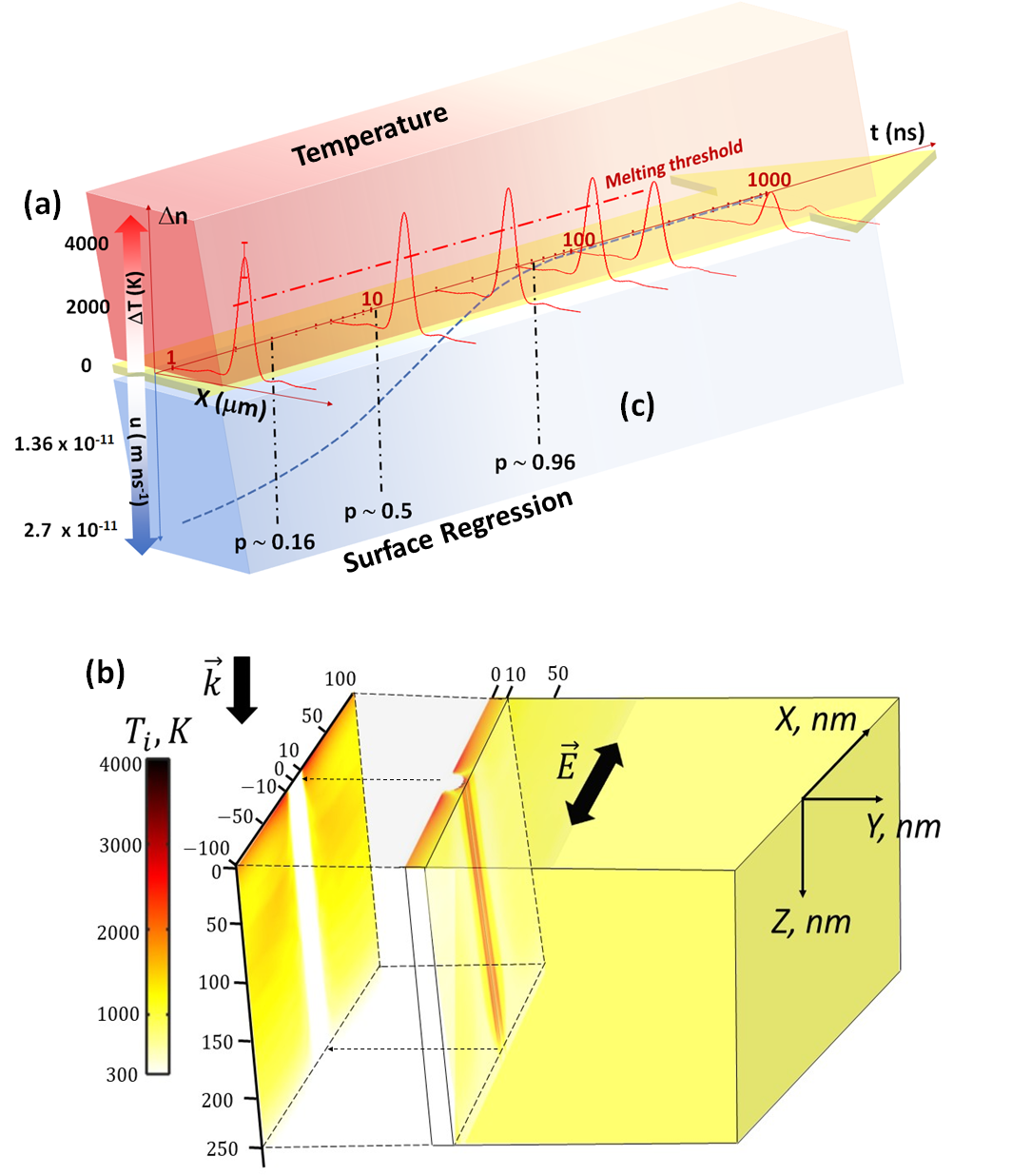}}
\caption{(a) Red curves illustrate temperature evolution from quantitative phase imaging of thermooptical changes in a single pulse (200\,fs) irradiation event in fused silica in the cavitation range. (b) Calculated field-driven temperature charts on silica interface $Z = 0$ with a hemispherical void nanotrench of $20$\,nm diameter centered at $(X, Y) = (0, 0)$. Two $XZ$ cross-sections are specified: at the interface $Y = 0$ on the left side and $Y = 10$ nm inside the bulk on the right side. The laser propagation $\vec{k}$ and laser polarization $\vec{E}$ are indicated. (c) Blue dashed line indicates the surface regression corresponding to the experimentally estimated temperatures, while dash-dotted black lines show the expected instantaneous number of removed monolayers of silica $p$.\label{Fig3}}
\end{figure*}

The nanoablation process puts forward a challenging mix of physical processes. Such highly confined irradiation conditions demand an accurate estimation of the thermodynamic state of the excited material as the subsequent transformation pathways in the restricted volume critically depend on the achievable temperatures and the cooling speed. The access to thermodynamic state properties, for example to the local temperature, can be achieved by monitoring temperature-dependent optical properties such as the refractive index via the thermooptic effect. Given the availability of thermooptical information, we test the thermodynamic evolution under relevant irradiation conditions for fused silica, a glass matching closely the optical and thermal characteristics of the AS\,87 aluminosilicate glass. The experiment is performed for a single pulse volume irradiation of fused silica using a similar Bessel-Gauss beam geometry, for irradiation conditions close but below the cavitation threshold \cite{somayaji20}. Ultrashort pulse excitation creates a thermal source with cylindrical symmetry, and thus a phase object that evolves in time according to the thermal transport. Using time-resolved quantitative phase microscopy we detect in real time the associated phase shift.  We thus observe with nanosecond time resolution the heat transport in the irradiation region in the fused silica sample, occurring, due to the cylindrical geometry, dominantly perpendicular on the optical axis. Observing the effect of the thermooptical evolution of the material \cite{Nguyen24,Raj25}, the phase and, respectively the refractive index detection (see in the Supporting Information) delivers peak values and spatial distribution of local temperatures. Acquired temperature data are given in Fig. ~\ref{Fig3} (a), indicating a local temperature exceeding 3000\,K in the first ns, with approximately 1\,$\mu$s decay time \cite{Bhu17} for a heat source of $0.75$\,$\mu$m diameter (full width half maximum). The irradiated nanoscale domain stays at a temperature between 3000\,K and 4000\,K for about 50\,ns. Given the below-threshold conditions for void formation, the temperature interval corresponds to a sustained molten region.

 Similar energetic considerations can be obtained by applying a theoretical model coupling the solution of Maxwell equations with the material topography and transient electronic and thermal response \cite{Zhang25} to calculate the field-enhanced energy deposition. The model is detailed in the Supporting Information. The results, in terms of field-driven temperature charts, are given in Fig. ~\ref{Fig3} (b), demonstrating how the energy is distributed in the vicinity of a void hemispheric nanotrench with $20$\,nm diameter, matching quite well the experimental data. In fact, the energy is absorbed efficiently at the tip of a hemispheric void trench on the sample interface in the scanning direction perpendicular laser polarization, corresponding to $Y$ axis. The maximum temperatures here reach $\approx 3500$\,K in sub-$10$\,nm area from the sample interface extending deep into the material. Overall, the background temperatures far from the void trench remain low $1500$\,K $- 2000$\,K to initiate any damage mechanisms. These data will allow identifying the nanoablation mechanisms.

\section{Discussion}
\subsection{Physical mechanisms of nanometer scale material removal}

Albeit the difficulties of grasping nanoscale properties of glass, the fundamental question is what ablative process is responsible for nm range material removal when, during the scan, multiple pulses (more than 100) illuminate a region comparable to the width. Material removal can be driven by different transformation mechanisms, including Hertz-Knudsen evaporation, nucleate boiling, phase explosion (spinodal decomposition), stress-confined spallation or fracture, nonthermal desorption and electrostatic ablation. We will discuss in the following the occurrence probability of several thermomechanical processes responsible for the observed nm ablation phase of glass driven by near-field cavity enhancement effects.

\textbf{Irradiation conditions} The field enhancement around a single nanohole $\mathrm{|\vec{E}|/E_0 \approx 3\epsilon_{glass}/(2+\epsilon_{glass}) \approx 1.59}$ ($\mathrm{\epsilon_{glass} = 2.25}$ is the AS\,87\, eco glass permittivity) in the poles perpendicular to laser polarization results in localized hot spots with temperatures exceeding twice the bulk temperatures upon laser excitation. As part of the thermal properties of the AS\,87\,eco aluminosilicate glass \cite{Schott} are less known, we will discuss for thermomechanical relaxation, as above, the generic case of fused silica, which may well approximate the present case. Specific estimates for AS 87 eco will be given whenever possible. For nanoscale ablation, close to the near-field ablation threshold, the maximum temperatures should be limited $T \leq 4000$ K. These localized temperatures persist laterally during $\mathrm{r^2/\kappa \approx 200}$\,ps with the diffusivity of silica $\mathrm{\kappa = k/\rho{C_i} \approx 5\cdot10^{-7}}$\,m$^2$/s ($\mathrm{T \approx 2000}$\,K) \cite{rudenko2018}, dimensions $\mathrm{r \approx 10}$\,nm. This estimate is valid for aluminosilicate glass having similar thermal diffusivity (thermal conductivity $\mathrm{k \approx 0.96}$ W/m/K, heat capacity $\mathrm{C_i = 840}$\,J/kg/K and the density $\mathrm{\rho \approx 2477}$\,kg/m$^3$). The amount of the ablated material by pulse is comparable then to the pitch of laser beam upon multipulse scanning (down to nanometer).

\textbf{Standard and non-standard mechanisms} Generally, spallation and phase explosion are the dominant processes that occur after intense ultrashort laser pulse irradiation. Faster than evaporation, the modification occurs when the system is still in strong thermodynamic non-equilibrium \cite{shugaev2021}. The material is removed in the form of clusters and nanodroplets, larger than a critical nuclei radius $r_{cr}$ in nucleate boiling, limited spall layer thickness in spallation or by the fastest growing wavelength in spinodal decomposition. Requiring larger volumes to develop, these mechanisms cannot explain the removal of a single monolayer $\approx 0.3$ nm when the temperatures are significantly lower than the critical value $\mathrm{T_{cr} \approx 5400}$\,K \cite{melosh2007}. For nucleation processes, according to classical nucleation theory (CNT), a critical nuclei radius $\mathrm{r_{cr} = - 2\sigma_0/{P_{cr}} \approx 3.17}$\,nm, where $\mathrm{P_{cr} = 189}$\,MPa is the critical pressure \cite{melosh2007} and $\mathrm{\sigma_0 = 0.3}$\,N/m \cite{boyd2012} is the surface tension, with a characteristic time $\mathrm{\tau_{nucl} = \exp(W/k_B{T})/N > 1}$\,s for $T <= 4000$ K, where $\mathrm{W = 2\pi{P_{cr}}{r}_{cr}^3/3}$ is the free energy at the critical radius and $\mathrm{N = V\rho{N_A}/{m_{SiO_2}} \approx 2208}$ is the number of available sites in $\mathrm{V = 10^3}$\,nm$^3$ volume. This estimate suggests that nucleation is improbable at the nanoscale (given the section of the affected volume) within a reasonable time and at sub-critical temperatures up to $\mathrm{T = 4000}$\,K.

Spallation can occur in solid or liquid phases. The characteristic time for spall in brittle solids, according to Grady \cite{grady1988}, $\mathrm{t_{spall} \approx \frac{1}{c_s}(\frac{6\sigma}{\rho\zeta^2})^{1/3} \approx 15}$\,ps and the characteristic spall size $\mathrm{l_{spall} \approx (\frac{48\sigma}{\rho\zeta^2})^{1/3} \approx 10}$\,nm, where $\mathrm{\zeta \approx {10}^{7}\frac{T}{T_{melt}}}$\,s$^{-1}$ is the thermal expansion-induced strain rate \cite{rudenko2018}, $\sigma = \sigma_0(1-T/T_{cr})^{1.25}$ is the surface tension, $T_{melt} = 1875$ K is the softening temperature of fused silica and $c_s = 3$\,km/s is the longitudinal sound velocity. This time is longer than electron-phonon coupling in silica $\approx 1$\,ps and transition to low viscosity glass, thus, the spallation is likely to occur in liquid phase instead. The scenario when the crack is already pre-formed by previous pulses and the probability to contribute to its growth in the solid phase after cooling is discussed further by applying the Griffith criterion. According to Grady's spall for liquid \cite{grady1988}, the sum of elastic and kinetic energies should be greater than the surface energy required to fracture the liquid into cavities of radius $R_{cav}$, plus the local viscous dissipation during void growth and coalescence in the cavitation process. According to the estimate for silica \cite{rudenko2018}, the cavitation in the bulk can occur for $T > 2000$\,K. The minimum time required for the spall to occur for $\mathrm{T = 3500-4000}$\,K is given mostly by the surface tension term $\mathrm{t \geq t_{min} = \sqrt{6\sigma/R_{cav}B\zeta^2}}$, where $\mathrm{B \approx 40}$\,GPa is the bulk modulus. Thus, for $\mathrm{R_{cav} = 1}$\,nm, $\mathrm{t_{min} \approx 6-10}$\,ns. For aluminosilicate glass, having one order higher thermal expansion and higher strain rates, the characteristic times are one order smaller, in the sub-nanosecond range.

It is also to be noted that the enhanced field of approximately $\mathrm{10^{10}}$\,V/m stays low compared to the requirements for charge separation at angstrom scales $\mathrm{|\vec{E}| < E_{crit} \approx e/{4\pi{r_{atom}}a\epsilon_0} \approx 3.6\cdot{10}^{10}}$\,V/m with a single silica layer $\mathrm{a = 0.3}$\,nm and atom radius $\mathrm{r \approx 0.13}$\,nm. An electrostatic repulsion (Coulomb explosion) is nonetheless essentially driven by residual charges and fields generated by electron photo- and thermionic emission \cite{Stoian2002} for times comparable with the laser pulse and with an action depth given by the nanometer range electronic mean free path \cite{Seah79} setting the extraction depth. Given the expected carrier densities in the range of the critical density at the incident wavelength $10^{22}$\,cm$^{-3}$ \cite{Bhu17}, such a condition may ensure sufficient charge deficit \cite{Bulgakova2004} and electrostatic fields of $\approx 10^{10}$\,V/m to trigger the particle ejection, assuming a quasi-complete depletion of the superficial layer. We estimate that such a process, concerning mainly charged particles (Si$^{+}$ and O$^{+}$) during the pulse duration, with an initial density comparable to the excitation density and an ionization rate of $10\%$, reflect a reduced fraction of the removed mass.

Hertz-Knudsen evaporation can occur in a system in near-equilibrium with quasi-steady temperatures and vapor pressure. Although these conditions are not satisfied at earlier times after ultrashort laser excitation, glass cooling takes a rather long time. The Hertz-Knudsen evaporation rate can be estimated as follows
\begin{equation}
\eta(T) \approx \frac{P_{sat}(T)}{\sqrt{2\pi{m_{SiO_2}}{k_B}T}},
\end{equation}
where $\mathrm{P_{sat}\approx 10^{A - B/T(K)}}$ is the temperature-dependent saturation vapor pressure expressed in bars, with $A = 8.203$ and $B = 25898.9$ \cite{melosh2007}. For $T = 3500$\,K, $\mathrm{P_{sat} \approx 6.4\cdot{10}^5}$\,Pa \cite{visscher2013}, the characteristic time of removal of a monolayer of area $a = 1$ nm$^2$, $\mathrm{\tau_{HK} = \frac{\sqrt{2\pi{m_{SiO_2}}RT}}{\alpha{N_A}{a}^2P_{sat}} \approx 2.76}$\,ns (here, $\mathrm{m_{SiO_2} = 60.2}$\,g/mole, $\mathrm{\alpha = 0.1}$ is the anomalous evaporation coefficient, $R$ and $N_A$ are gas constant and Avogadro numbers), and, of one silica layer of $\approx 0.3$\,nm, $\mathrm{\tau_{HK} \approx 30.7}$\,ns. For $T = 4000$\,K, $\mathrm{P_{sat} \approx 5.3\cdot{10}^6}$\,Pa \cite{visscher2013}, the characteristic time of removal ${a}^2 = 1^2$\,nm$^2$ area, $\mathrm{\tau_{HK} = \frac{\sqrt{2\pi{m_{SiO_2}}RT}}{\alpha{N_A}{a}^2P_{sat}} \approx 0.39}$\,ns and, of one silica layer of $\approx 0.3$\,nm, $\tau_{HK} \approx 3.5$\,ns. Within these characteristic times, the temperature and the vapor pressure remain steady and the system can be considered in near-equilibrium.

These times are consistent with quantitative temperature measurements in Fig. \ref{Fig3}(a) and simulation results in Fig. \ref{Fig3}(b). We indicate the corresponding surface regression velocities $u = \eta(T) / n_a$ by a blue dashed line in Fig. ~\ref{Fig3} (c), where atomic number density for fused silica is defined as $\mathrm{n_a = \rho*N_A/m_{SiO_2} \approx 2.2{\cdot}10^{28} \ m^{-3}}$. The velocities are in the range of $\mathrm{10^{-11} - 10^{-10} \ m/ns}$, resulting in a nanometer range ablative phenomenon, and decay exponentially several tens of nanoseconds after laser irradiation. These values are sufficient to remove eventually one monolayer of glass at sub-$100$ ns time scales before the lattice cools down as shown in Fig. ~\ref{Fig3} (c), where $p$ stands for the expected number of removed monolayers, integrating the dynamic velocities by time delay. The velocities stay similar if we consider the highly probable non-congruent ablation in the form of decomposed silica and $SiO$ and $O_2$ vapor components. It should be noted that even if these temperatures and negative pressures commonly induce cavitation and spallation after ultrashort laser surface irradiation, the spall layer thickness is typically larger than few nanometers.

From a mechanical viewpoint, the Griffith criterion \cite{Griffithcriterion} predicts the fracture stress required to apply to a crack for spall growth as a function of crack size as follows $\mathrm{\Sigma = \sqrt{2Y\gamma/\pi{a}}}$, where $\mathrm{Y \approx 72}$\, GPa is the Young's modulus, $\mathrm{\gamma = 1}$\,J/m$^2$ is the surface energy density. For $\mathrm{a = 1-10}$\,nm, the required stress is $\mathrm{\Sigma_{cr} \approx 2-7}$\,GPa. The tensile thermal stress upon cooling is much lower $\mathrm{\Sigma \approx Y\alpha_T\Delta{T} \approx 80}$\,MPa for $\mathrm{\Delta{T} \approx 2000}$\,K and $\mathrm{\alpha_T = 5.5\cdot{10}^{-7}}$\,K$^{-1}$ for fused silica \cite{kikuchi1997} and $\mathrm{\Sigma \approx 1.27}$\,GPa with $\mathrm{\alpha_T = 8.74\cdot{10}^{-6}}$\,K$^{-1}$ for aluminosilicate glass. However, in a stress-confined regime, continuum elasticity breaks down, larger stresses can be induced and they scale down with the size. The upper estimate is given by $\mathrm{\Sigma \approx \rho{c_s}^2\alpha_T\Delta{T}(L/L_{cr}) \leq 8.7}$\,GPa for fused silica and $\Sigma \leq 154.9$\,GPa for aluminosilicate glass. Here, we assume $L_{cr} = 10$\,nm is the critical size and the longitudinal sound velocity $\mathrm{c_s = 3-6}$\,km/s, depending on the temperature \cite{vo2005}. As a result, the steady conditions for crack propagation are equally relevant at the nanoscale.

The estimations above indicate thus as most probable nanodicing mechanism a mix of nanofracturing and Hertz-Kundsen evaporation of the walls. Notwithstanding, the elastic properties of nanoscale silica and the rapid achievement of low viscosity states may arrest the process. Rapid heating and the development of melts will favor cavitation with a potential increase of the processing scale to several tens of nanometer. Such energy density dependence matches well the differences observed between fs and ps pulse irradiation, where the latter increases the deposited energy density by better nonlinear stability, avoiding plasma defocusing \cite{Bhuyan2014}

All the equilibrium thermal phase transformations discussed here may have non-equilibrium equivalents on the timescale of the pulse duration. These are the result of the modification of interatomic potentials due to electronic excitation and entropic effects on the free energy of the system \cite{Ben21}.

\section{Conclusion}
We have demonstrated a direct, near-field-enhanced, non-diffractive ultrafast laser cutting process that yields glass surface with nanometer-scale surface roughness, comparable to optically polishing. Combining ultrafast Bessel-Gauss beams with near-field energy localization confines the material modification to several molecular layers, suppressing chipping, and catastrophic subsurface damage that typically limit laser cutting. Separation proceeds through nanofracture, while wall recession is dominated by Hertz-Knudsen evaporation, resulting in nanometer-scale surface regression. These results show that glass cutting can be effectively transformed into an extreme nanometer-scale cutting tool with an interface quality previously attainable only through complex polishing. This opens new opportunities for integrating glass into high-performance optical, microfluidic, and nanoscale devices, offering a route to extend near-field-enabled laser processing strategies to other brittle transparent materials.

\section{\label{Method} Materials and methods}

\subsection{Materials}
Thin Schott AS\,87\,eco aluminosilicate glass layers ($\mathrm{30\times20\times0.1}$\,mm$^3$) were laser irradiated. The glass is characterized by shock resistance, large thermomechanical expansion, and a transparency window, exceeding 250\,nm. The direction of the beam is perpendicular to the polished surface, the polarization is set perpendicular to the scan direction and the pitch was set in the range down to 1\,nm.

\subsection{Experimental setup}
Ultrafast non-diffractive $\mathrm{0^{th}}$ order Bessel-Gauss pulses were generated by illuminating an axicon (apex angle of $\mathrm{178}^{\circ}$) with a Gaussian-profiled laser beam from an ultrafast laser system operating at a central wavelength of $\mathrm {1.03}$\,$\mu$m, with a maximum average power of 6\,W, a tunable repetition rate $\mathrm{50- 600}$\,kHz, and adjustable pulse duration from $\mathrm{200}$\,fs to $\mathrm{5}$ \,ps. A tunable beam expander (1$\times-$ 4$\times$) was employed to control the beam diameter incident on the beam-shaping optics and thus the axial projection of the Bessel-Gauss beam. A 4-f imaging system, comprising a $\mathrm{500}$\,mm focal-length lens and a $\mathrm{20\times}$ infinity-corrected Mitutoyo objective lens (effective focal length $\mathrm{10}$\,mm and $\mathrm{0.42 \ NA}$), was used to relay and demagnify the Bessel beam by a factor of $\mathrm{50}$ in the sample. The conical wavefronts have a half-cone angle of  $\mathrm{20^{\circ}}$ in free space. The ultrafast Bessel pulse featured a central core diameter of ~$\mathrm{1.1}$\,$\mu$m FWHM and a Bessel length of ~$\mathrm{120}$\,$\mu$m in air. The sample was mounted on a three-axis nanopositioning platform, enabling precise spatial manipulation during experiments. The beam was designed so that the peak intensity compensates for the expected drop as the cone angle decreases with penetration into the glass, ensuring a constant intensity profile inside the glass, slightly above the threshold for surface ablation. The laser process was mostly carried out at a nominal repetition rate of $\mathrm{100}$\,kHz for a scan speed down to $\mathrm{0.1}$\,mm/s, corresponding to a lowest pitch of $\mathrm{1}$\,nm (further details in Supporting Information).

\subsection{Temperature estimation}
The temperature evolution was determined by time-resolved quantitative phase microscopy of the laser-heated regions. Single laser pulses were used for irradiation. The optical phase shift at a specific time moment is obtained by an experimental setup described in \cite{Raj25}, involving here as the irradiation source a laser pulse at $\mathrm {\lambda = 1.03}$\,$\mu$m and as the observation source a random laser pulse of 7\,ns duration and 590\,nm wavelength. The method, based on common path digital holography \cite{Pop14}, detects transient phase objects associated with temperature upon laser irradiation of fused silica, providing absolute phase-shift cartography. Transient refractive index data are extracted from the inverse Abel transform of the optical phase chart and deconvoluted using the characteristic point spread function of the microscope (550\,nm optical resolution). The temperatures are extracted from the thermooptical evolution of the refractive index $\Delta n=\partial{n} / \partial{T} \Delta T$ with a thermooptical coefficient of $\mathrm {8.74 \cdot {10}^{-6}}$\,K $^{-1}$ \cite{rego2023temperature}.

\subsection{Model}
We apply a multi-physical model to estimate the temperatures resulted from ultrashort laser excitation with laser irradiation parameters similar to experimental. The model was previously described, and material parameters for fused silica were adopted from \cite{Rud21, Zhang25}. A Finite-Difference Time-Domain (FDTD) approach is applied to solve the system of Maxwell's equations in the vicinity of the scatterer and at a laser ablation fluence above threshold. The nonlinear Maxwell equations \cite{rudenko2018} were coupled with the rate equation for free carrier generation and electron and ion temperature dynamics (two-temperature model) \cite{Rud21}. The temperature dynamics is resolved for 10 ps until thermal equilibrium is reached and the maximum ion temperatures are established. The input optical source is a Bessel-shaped laser pulse with $\theta$ = 14$^\circ$ in glass, the peak fluence 3\,J/cm$^2$, pulse duration of 200 fs and central laser wavelength of 1030 nm. The pulse duration is considered generic, and the use of other pulse durations will only qualitatively influence the thermal levels. Simulation implies the initial infinite hemispheric void trench on silica interface with a 20 nm-diameter centred at $\mathrm{[X, Y]=(0,0)}$ nm, $X$ axis being along the laser polarization. The ultrashort laser source excitation along the Z axis is crossing the sample at Z = 0, focused on the void structure. The whole simulation domain is $800{\times}800{\times}400$\,nm$^3$, having $\mathrm{\Delta{X,Y,Z} = 1}$\,nm resolution.

\subsection{Characterization}
Temperature estimates are obtained using quantitative phase-contrast microscopy and the thermo-optical coefficient of the sample (see supporting information).

Scanning electron microscopy (SEM) was employed to characterize both the surface morphology of the laser-written nanolines and the laser-separated surfaces. Atomic force microscopy (AFM) was used to quantitatively analyze the surface topography and roughness of the laser-fabricated nanostructures.

%\backmatter
\subsection*{Author Contributions}

This is an author contribution text. This is an author contribution text. This is an author contribution text. This is an author contribution text. This is an author contribution text.

\subsection*{Acknowledgments}
We would like to thank the Banque Publique D'Investissement France (iDEMO grant Glacier) and the French National Research Agency (grant ANR-21-CE08-0005) for their financial support. We also acknowledge V. Maffeis (IREIS) and A. Cazier and N. Faure for their support.

\subsection*{Conflicts of Interest}

The authors declare no conflicts of interest.

\subsection*{Data Availability Statement}

This is a data availability statement.

\bibliography{references}

@article{gattass2008femtosecond,
  title={Femtosecond laser micromachining in transparent materials},
  author={Gattass, Rafael R and Mazur, Eric},
  journal={Nat. photonics},
  volume={2},
  number={4},
  pages={219--225},
  year={2008},
  publisher={Nature Publishing Group UK London}
}

@article{Bhuyan2014,
author = {Bhuyan,M. K.  and Velpula,P. K.  and Colombier,J. P.  and Olivier,T.  and Faure,N.  and Stoian,R. },
title = {Single-shot high aspect ratio bulk nanostructuring of fused silica using chirp-controlled ultrafast laser Bessel beams},
journal = {Appl. Phys. Lett.},
volume = {104},
number = {2},
pages = {021107},
year = {2014},
doi = {10.1063/1.4861899},
}

@article{somayaji20,
  TITLE = {{Multiscale electronic and thermomechanical dynamics in ultrafast nanoscale laser structuring of bulk fused silica}},
  AUTHOR = {Somayaji, Madhura and Bhuyan, Manoj and Bourquard, Florent and Velpula, Praveen and D'amico, Ciro and Colombier, Jean-Philippe and Stoian, Razvan},
  JOURNAL = {{Scientific Reports}},
  PUBLISHER = {{Nature Publishing Group}},
  VOLUME = {10},
  NUMBER = {1},
  YEAR = {2020},
  MONTH = Dec,
  DOI = {10.1038/s41598-020-71819-9},
}

@article{Courvoisier:09,
author = {F. Courvoisier and P.-A. Lacourt and M. Jacquot and M. K. Bhuyan and L. Furfaro and J. M. Dudley},
journal = {Opt. Lett.},
number = {20},
pages = {3163--3165},
publisher = {Optica Publishing Group},
title = {Surface nanoprocessing with nondiffracting femtosecond Bessel beams},
volume = {34},
month = {Oct},
year = {2009},
doi = {10.1364/OL.34.003163},
}

@article{bhuyan2015high,
  title={High-speed laser-assisted cutting of strong transparent materials using picosecond Bessel beams},
  author={Bhuyan, MK and Jedrkiewicz, O and Sabonis, V and Mikutis, M and Recchia, Sandro and Aprea, A and Bollani, M and Trapani, P Di},
  journal={Appl. Phys. A},
  volume={120},
  number={2},
  pages={443--446},
  year={2015},
  publisher={Springer}
}

@article{Jenne:20,
author = {Michael Jenne and Daniel Flamm and Keyou Chen and Marcel Sch\"{a}fer and Malte Kumkar and Stefan Nolte},
journal = {Opt. Express},
number = {5},
pages = {6552--6564},
publisher = {Optica Publishing Group},
title = {Facilitated glass separation by asymmetric Bessel-like beams},
volume = {28},
year = {2020},
doi = {10.1364/OE.387545},
}

@misc{chen2019,
      title={Generalized axicon-based generation of nondiffracting beams},
      author={Keyou Chen and Michael Jenne and Daniel GÃ¼nther Grossmann and Daniel Flamm},
      year={2019},
      eprint={1911.03103},
      archivePrefix={arXiv},
      primaryClass={physics.optics},
      url={https://arxiv.org/abs/1911.03103},
}

@article{Dudutis:20,
author = {Juozas Dudutis and Jok\={u}bas Pipiras and Rokas Stonys and Eimantas Daknys and Art\={u}ras Kilikevi\v{c}ius and Albinas Kasparaitis and Gediminas Ra\v{c}iukaitis and Paulius Ge\v{c}ys},
journal = {Opt. Express},
number = {21},
pages = {32133--32151},
publisher = {Optica Publishing Group},
title = {In-depth comparison of conventional glass cutting technologies with laser-based methods by volumetric scribing using Bessel beam and rear-side machining},
volume = {28},
year = {2020},
doi = {10.1364/OE.402567},
}

@article{Bhu10,
    author = {Bhuyan, M. K. and Courvoisier, F. and Lacourt, P. A. and Jacquot, M. and Salut, R. and Furfaro, L. and Dudley, J. M.},
    title = {High aspect ratio nanochannel machining using single shot femtosecond Bessel beams},
    journal = {Appl. Phys. Lett.},
    volume = {97},
    number = {8},
    pages = {081102},
    year = {2010},
    doi = {10.1063/1.3479419}
}

@article{Bhu17,
author = {M. K. Bhuyan and M. Somayaji and A. Mermillod-Blondin and F. Bourquard and J. P. Colombier and R. Stoian},
journal = {Optica},
number = {8},
pages = {951--958},
title = {Ultrafast laser nanostructuring in bulk silica, a ``slow'' microexplosion},
volume = {4},
year = {2017},
doi = {10.1364/OPTICA.4.000951}
}

@article{shin2020strength,
  title={Strength of ultra-thin glass cut by internal scribing using a femtosecond Bessel beam},
  author={Shin, Hyesung and Kim, Dongsik},
  journal={Opt. Laser Technol.},
  volume={129},
  pages={106307},
  year={2020},
  publisher={Elsevier}
}

@article{meyer2017submicron,
  title={Submicron-quality cleaving of glass with elliptical ultrafast Bessel beams},
  author={Meyer, R{\'e}mi and Giust, Remo and Jacquot, Maxime and Dudley, John Micha{\"e}l and Courvoisier, Fran{\c{c}}ois},
  journal={Appl. Phys. Lett.},
  volume={111},
  number={23},
  year={2017},
  publisher={AIP Publishing}
}

@article{sugioka2014ultrafast,
  title={Ultrafast lasers - reliable tools for advanced materials processing},
  author={Sugioka, Koji and Cheng, Ya},
  journal={Light: Sci. Appl.},
  volume={3},
  number={4},
  pages={e149--e149},
  year={2014},
  publisher={Nature Publishing Group}
}

@article{malinauskas2016ultrafast,
  title={Ultrafast laser processing of materials: from science to industry},
  author={Malinauskas, Mangirdas and {\v{Z}}ukauskas, Albertas and Hasegawa, Satoshi and Hayasaki, Yoshio and Mizeikis, Vygantas and Buividas, Ri{\v{c}}ardas and Juodkazis, Saulius},
  journal={Light: Sci. Appl.},
  volume={5},
  number={8},
  pages={e16133--e16133},
  year={2016},
  publisher={Nature Publishing Group}
}

@article{rapp2017high,
  title={High speed cleaving of crystals with ultrafast Bessel beams},
  author={Rapp, Ludovic and Meyer, R{\'e}mi and Furfaro, Luca and Billet, Cyril and Giust, R and Courvoisier, F},
  journal={Opt. Express},
  volume={25},
  number={8},
  pages={9312--9317},
  year={2017},
  publisher={Optical Society of America}
}

@article{ungaro2021using,
  title={Using phase-corrected Bessel beams to cut glass substrates with a chamfered edge},
  author={Ungaro, Craig and Kaliteevskiy, Nikolay and Sterlingov, Petr and Ivanov, Viacheslav V and Boh Ruffin, A and Terbrueggen, Ralf J and Savidis, Nickolaos},
  journal={Appl. Opt.},
  volume={60},
  number={3},
  pages={714--719},
  year={2021},
  publisher={Optical Society of America}
}

@article{balage2023bessel,
  title={Bessel beam dielectrics cutting with femtosecond laser in GHz-burst mode},
  author={Balage, Pierre and Guilberteau, Th{\'e}o and Lafargue, Manon and Bonamis, Guillaume and H{\"o}nninger, Clemens and Lopez, John and Manek-H{\"o}nninger, Inka},
  journal={Micromachines},
  volume={14},
  number={9},
  pages={1650},
  year={2023},
  publisher={MDPI}
}

@article{liu2024high,
  title={High-Quality Cutting of Soda--Lime Glass with Bessel Beam Picosecond Laser: Optimization of Processing Point Spacing, Incident Power, and Burst Mode},
  author={Liu, Jiaxuan and Yang, Jianjun and Chen, Hui and Li, Jinxuan and Zhang, Decheng and Zhong, Jian and Pan, Xinjian},
  journal={Appl. Sci.},
  volume={14},
  number={5},
  pages={1885},
  year={2024},
  publisher={MDPI}
}

@article{dudutis2025polarization,
  title={Polarization-Dependent Laser-Assisted Cutting of Glass Using a Nondiffractive Beam in the MHz Burst Regime},
  author={Dudutis, Juozas and Kondratas, Aleksandras and Gecys, Paulius},
  journal={ACS Photonics},
  year={2025},
  publisher={ACS Publications}
}

@article{balage2025pump,
  title={Pump-Probe Imaging of Ultrafast Laser Percussion Drilling of Glass in Single Pulse, MHz-and GHz-Burst Regimes},
  author={Balage, Pierre and Guilberteau, Th{\'e}o and Lafargue, Manon and Bonamis, Guillaume and H{\"o}nninger, Clemens and Lopez, John and Manek-H{\"o}nninger, Inka},
  journal={Adv. Mater. Interf},
  volume={12},
  number={10},
  pages={2400853},
  year={2025},
  publisher={Wiley Online Library}
}

@article{Zhang25,
author = {Guodong Zhang  and Anton Rudenko  and Razvan Stoian  and Guanghua Cheng },
title = {Ultrafast Laser High-Aspect-Ratio Extreme Nanostructuring of Glass beyond $\lambda/100$},
journal = {Ultrafast sci.},
volume = {5},
number = {},
pages = {0103},
year = {2025},
doi = {10.34133/ultrafastscience.0103},
}

@article{li2024super,
  title={Super-stealth dicing of transparent solids with nanometric precision},
  author={Li, Zhen-Ze and Fan, Hua and Wang, Lei and Zhang, Xu and Zhao, Xin-Jing and Yu, Yan-Hao and Xu, Yi-Shi and Wang, Yi and Wang, Xiao-Jie and Juodkazis, Saulius and others},
  journal={Nat. Photonics},
  volume={18},
  number={8},
  pages={799--808},
  year={2024},
  publisher={Nature Publishing Group UK London}
}

@article{yan2022near,
  title={Near-Field Mediated 40 nm In-Volume Glass Fabrication by Femtosecond Laser},
  author={Yan, Zhi and Gao, Jichao and Beresna, Martynas and Zhang, Jingyu},
  journal={Adv. Opt. Mater.},
  volume={10},
  number={4},
  pages={2101676},
  year={2022},
  publisher={Wiley Online Library}
}

@article{liu2023engraving,
  title={Engraving depth-controlled nanohole arrays on fused silica by direct short-pulse laser ablation},
  author={Liu, Xin and Clady, Rapha{\"e}l and Grojo, David and Ut{\'e}za, Olivier and Sanner, Nicolas},
  journal={Adv. Mater. Interf.},
  volume={10},
  number={7},
  pages={2202189},
  year={2023},
  publisher={Wiley Online Library}
}

@article {Mie1908,
author = {Mie, Gustav},
title = {Beitr\"{a}ge zur Optik tr\"{u}ber Medien, speziell kolloidaler Metall\"{o}sungen},
journal = {Ann. Phys.},
volume = {330},
number = {3},
publisher = {WILEY-VCH Verlag},
issn = {1521-3889},
url = {http://dx.doi.org/10.1002/andp.19083300302},
doi = {10.1002/andp.19083300302},
pages = {377--445},
year = {1908},
}

@article{kikuchi1997,
  title={Thermal expansion of vitreous silica: Correspondence between dilatation curve and phase transitions in crystalline silica},
  author={Kikuchi, Yoshikazu and Sudo, Hajime and Kuzuu, Nobu},
  journal={J. Appl. Phys.},
  volume={82},
  number={8},
  pages={4121--4123},
  year={1997},
  publisher={American Institute of Physics}
}

@article{kelly1999,
  title={Contribution of vaporization and boiling to thermal-spike sputtering by ions or laser pulses},
  author={Kelly, Roger and Miotello, Antonio},
  journal={Phys. Rev. E},
  volume={60},
  number={3},
  pages={2616},
  year={1999},
  publisher={APS}
}

@article{kelly2000,
  title={Does normal boiling exist due to laser-pulse or ion bombardment?},
  author={Kelly, Roger and Miotello, Antonio},
  journal={J. Appl. Phys.},
  volume={87},
  number={6},
  pages={3177--3179},
  year={2000},
  publisher={American Institute of Physics}
}

@article{rudenko2018,
  title={Nanopore-mediated ultrashort laser-induced formation and erasure of volume nanogratings in glass},
  author={Rudenko, Anton and Colombier, Jean-Philippe and Itina, Tatiana E},
  journal={Phys. Chem. Chem. Phys.},
  volume={20},
  number={8},
  pages={5887--5899},
  year={2018},
  publisher={Royal Society of Chemistry}
}

@book{shugaev2021,
  title={Laser-induced thermal processes: heat transfer, generation of stresses, melting and solidification, vaporization, and phase explosion},
  author={Shugaev, Maxim V and He, Miao and Levy, Yoann and Mazzi, Alberto and Miotello, Antonio and Bulgakova, Nadezhda M and Zhigilei, Leonid V},
  booktitle={Handbook of Laser Micro-and Nano-Engineering},
  pages={83--163},
  year={2021},
  publisher={Springer}
}

@article{melosh2007,
  title={A hydrocode equation of state for SiO2},
  author={Melosh, HJ},
  journal={Meteorit. Planet. Sci.},
  volume={42},
  number={12},
  pages={2079--2098},
  year={2007},
  publisher={Wiley Online Library}
}

@article{elhadj2012,
  title={Evaporation kinetics of laser heated silica in reactive and inert gases based on near-equilibrium dynamics},
  author={Elhadj, Selim and Matthews, Manyalibo J and Yang, Steven T and Cooke, Diane J},
  journal={Opt. Express},
  volume={20},
  number={2},
  pages={1575--1587},
  year={2012},
  publisher={Optical Society of America}
}

@article{ma2025,
  title={Multiscale study of laser-induced dynamic fracture: phenomena, mechanisms and applications},
  author={Ma, Yiming and Li, Fang and Tian, Hong and Zuo, Pei and Zhou, Yaowei},
  journal={Eng. Fract. Mech.},
  pages={111687},
  year={2025},
  publisher={Elsevier}
}

@article{kraus2012,
  title={Shock vaporization of silica and the thermodynamics of planetary impact events},
  author={Swift, DC and Bolme, CA and Smith, RF and Hamel, S and Hammel, BD and Spaulding, DK and Hicks, DG and Eggert, JH and others},
  journal={J. Geophys. Res. Planets},
  volume={117},
  number={E9},
  year={2012},
  publisher={Wiley Online Library}
}

@article{visscher2013,
  title={Chemistry of impact-generated silicate melt-vapor debris disks},
  author={Visscher, Channon and Fegley, Bruce},
  journal={Astrophys. J. Lett.},
  volume={767},
  number={1},
  pages={L12},
  year={2013},
  publisher={IOP Publishing}
}

@phdthesis{bercegol2009,
  title={Endommagement laser nanoseconde en surface de la silice vitreuse},
  author={Bercegol, Herv{\'e}},
  year={2009},
  school={Universit{\'e} Bordeaux 1}
}

@article{vo2005,
  title={Sound velocity in alumino-silicate liquids determined up to 2550 K from Brillouin spectroscopy: Glass transition and crossover temperatures},
  author={Vo-Thanh, Dung and Bottinga, Yan and Polian, Alain and Richet, Pascal},
  journal={J. Non-Cryst. Solids},
  volume={351},
  number={1},
  pages={61--68},
  year={2005},
  publisher={Elsevier}
}

@online{Schott,
  author = {http://https://www.schott.com/fr-fr/products/as-87-neo-p1000312}
}

@article{Bulgakova2004,
  title = {Electronic transport and consequences for material removal in ultrafast pulsed laser ablation of materials},
  author = {Bulgakova, N. M. and Stoian, R. and Rosenfeld, A. and Hertel, I. V. and Campbell, E. E. B.},
  journal = {Phys. Rev. B},
  volume = {69},
  issue = {5},
  pages = {054102},
  numpages = {12},
  year = {2004},
  month = {Feb},
  publisher = {American Physical Society},
  doi = {10.1103/PhysRevB.69.054102},
}

@article{Stoian2002,
  title = {Coulomb explosion in ultrashort pulsed laser ablation of ${\mathrm{Al}}_{2}{\mathrm{O}}_{3}$},
  author = {Stoian, R. and Ashkenasi, D. and Rosenfeld, A. and Campbell, E. E. B.},
  journal = {Phys. Rev. B},
  volume = {62},
  issue = {19},
  pages = {13167--13173},
  numpages = {0},
  year = {2000},
  month = {Nov},
  publisher = {American Physical Society},
  doi = {10.1103/PhysRevB.62.13167},
}

@article{Seah79,
author = {Seah, M. P. and Dench, W. A.},
title = {Quantitative electron spectroscopy of surfaces: A standard data base for electron inelastic mean free paths in solids},
journal = { Surf. Interface Anal.},
volume = {1},
number = {1},
pages = {2-11},
doi = {https://doi.org/10.1002/sia.740010103},
year = {1979}
}

@article{Rud21,
author = {Rudenko, Anton and Colombier, Jean-Philippe and Itina, Tatiana E. and Stoian, Razvan},
title = {Genesis of nanogratings in silica bulk via multipulse interplay of ultrafast photo-excitation and hydrodynamics},
journal = {Adv. Opt. Mater.},
volume = {9},
number = {20},
pages = {2100973},
doi = {10.1002/adom.202100973},
year = {2021}
}

@article{Ben21,
  title = {Structural instability of transition metals upon ultrafast laser irradiation},
  author = {Ben-Mahfoud, L. and Silaeva, E. P. and Stoian, R. and Colombier, J. P.},
  journal = {Phys. Rev. B},
  volume = {104},
  issue = {10},
  pages = {104104},
  numpages = {10},
  year = {2021},
  publisher = {American Physical Society},
  doi = {10.1103/PhysRevB.104.104104},
}

@article{grady1988,
  title={The spall strength of condensed matter},
  author={Grady, Dennis E},
  journal={J. Mech. Phys.},
  volume={36},
  number={3},
  pages={353--384},
  year={1988},
  publisher={Elsevier}
}

@article{Raj25,
author = {Dwivedi, Rajeev and Nguyen, Huu Dat and Joao, Sergio Sao and Seydoux-Guillaume, Anne-Magali and Kuppan, Thirunaukkarasu and D'Amico, Ciro and Kermouche, Guillaume and Stoian, Razvan},
title = {Dynamic Ultrafast Laser-Induced Structural Changes and Extreme Nanostructuring in Hard Dielectric Materials},
journal = {ACS Photonics},
volume = {12},
number = {7},
pages = {3644-3652},
year = {2025},
doi = {10.1021/acsphotonics.5c00554},
URL = {https://doi.org/10.1021/acsphotonics.5c00554},
}

@article{rego2023temperature,
  title={Temperature dependence of the thermooptic coefficient of SiO$_2$ glass},
  author={Rego, Gaspar},
  journal={Sensors},
  volume={23},
  number={13},
  pages={6023},
  year={2023},
  publisher={MDPI}
}

@article{boyd2012,
author = {Keiron Boyd and Heike Ebendorff-Heidepriem and Tanya M. Monro and Jesper Munch},
journal = {Opt. Mater. Express},
number = {8},
pages = {1101--1110},
publisher = {Optica Publishing Group},
title = {Surface tension and viscosity measurement of optical glasses using a scanning CO$_2$ laser},
volume = {2},
year = {2012},
doi = {10.1364/OME.2.001101}
}

@article{Nguyen24,
author = {H. D. Nguyen  and A. Tsaturyan  and S. Sao Joao  and R. Dwivedi  and A. Melkonyan  and C. D'Amico  and E. Kachan  and J. P. Colombier  and G. Kermouche  and R. Stoian },
title = {Quantitative Mapping of Transient Thermodynamic States in Ultrafast Laser Nanostructuring of Quartz},
journal = {Ultrafast Sci.},
volume = {4},
number = {},
pages = {0056},
year = {2024},
doi = {10.34133/ultrafastscience.0056},
URL = {https://spj.science.org/doi/abs/10.34133/ultrafastscience.0056},
}

@ARTICLE{Pop14,
  author = {Bhaduri, B. and Edwards, C. and Pham, H. and Zhou, R. and Nguyen, T. H. and Goddard, L. L. and Popescu, G.},
  title = {Diffraction phase microscopy: principles and applications in materials and life sciences},
  journal = {Adv. Opt. Photonics.},
  volume = {6},
  pages = {57},
  year = {2014}
}

@article{nanofludic,
author = {Osellame, R. and Hoekstra, H.J.W.M. and Cerullo, G. and Pollnau, M.},
title = {Femtosecond laser microstructuring: an enabling tool for optofluidic lab-on-chips},
journal = {Laser Photonics Rev.},
volume = {5},
number = {3},
pages = {442-463},
doi = {https://doi.org/10.1002/lpor.201000031},
year = {2011}
}

@article{photonicssystemFAN20088,
title = {Sensitive optical biosensors for unlabeled targets: A review},
journal = {Anal. Chim. Acta},
volume = {620},
number = {1},
pages = {8-26},
year = {2008},
issn = {0003-2670},
doi = {https://doi.org/10.1016/j.aca.2008.05.022},
author = {Xudong Fan and Ian M. White and Siyka I. Shopova and Hongying Zhu and Jonathan D. Suter and Yuze Sun},
}

@ARTICLE{fibre-to-chip-waveguide,
    
AUTHOR={Grenier, Jason R.  and Brusberg, Lars  and Wieland, Kristopher A.  and Matthies, Juergen  and Terwilliger, Chad C. },
           
TITLE={Ultrafast laser processing of glass waveguide substrates for multi-fiber connectivity in co-packaged optics},
          
JOURNAL={Adv. Opt. Technol.},
          
VOLUME={Volume 12 - 2023},
  
YEAR={2023},
  
DOI={10.3389/aot.2023.1244009},
  
ISSN={2192-8584},
}

@article{Fluence,
    author = {Lamperti, M. and Jukna, V. and Jedrkiewicz, O. and Di Trapani, P. and Stoian, R. and Itina, T. E. and Xie, C. and Courvoisier, F. and Couairon, A.},
    title = {Invited Article: Filamentary deposition of laser energy in glasses with Bessel beams},
    journal = {APL Photon.},
    volume = {3},
    number = {12},
    pages = {120805},
    year = {2018},
    month = {12},
    issn = {2378-0967},
    doi = {10.1063/1.5053085},
}

@article{Griffithcriterion,
author = {Griffith, Alan Arnold},
    title = {VI. The phenomena of rupture and flow in solids},
    journal = {Philos. Trans. A. Math. Phys. Eng. Sci.},
    volume = {221},
    number = {582-593},
    pages = {163-198},
    year = {1921},
    month = {01},
    issn = {0264-3952},
    doi = {10.1098/rsta.1921.0006},
}

\nocite{*}% Show all bib entries - both cited and uncited; comment this line to view only cited bib entries;

\end{document}